\pdfoutput=1
\documentclass[aps,prl,twocolumn,superscriptaddress,amsmath,amssymb,floatfix,longbibliography]{revtex4-2}

\usepackage{graphicx}
\usepackage{bm}
\usepackage{xcolor}
\usepackage{hyperref}
\hypersetup{colorlinks=true,linkcolor=blue!70!black,citecolor=magenta!75!black,urlcolor=blue!70!black,
  pdftitle={Can One Hear the Size of a Clock?},pdfauthor={Markus Baumann and Claudia Linnhoff-Popien}}
\usepackage{orcidlink}
\usepackage{amsthm}
\newtheorem{theorem}{Theorem}
\newtheorem{lemma}{Lemma}
\newtheorem{corollary}{Corollary}

\newcommand{\arccot}{\operatorname{arccot}}
\newcommand{\Ft}{F_{\mathrm{tot}}}
\newcommand{\runin}[1]{\textit{\textbf{#1}}---}
\newcommand{\SM}{\cite{SM}}
\newcommand{\SMsec}[1]{\cite[Sec.~\ref{#1}]{SM}}

\begin{document}

\title{Can One Hear the Size of a Clock?}

\author{Markus Baumann\,\orcidlink{0009-0007-3575-1006}}
\email[Contact author: ]{markus.baumann@campus.lmu.de}
\affiliation{Quantum Applications and Research Laboratory (QAR-Lab),\\ Ludwig-Maximilians-Universit\"at M\"unchen, Oettingenstra{\ss}e 67, 80538 M\"unchen, Germany}
\author{Claudia Linnhoff-Popien\,\orcidlink{0000-0001-6284-9286}}
\affiliation{Quantum Applications and Research Laboratory (QAR-Lab),\\ Ludwig-Maximilians-Universit\"at M\"unchen, Oettingenstra{\ss}e 67, 80538 M\"unchen, Germany}

\date{November 6, 2026}

\begin{abstract}
\looseness=-1 Weyl's law counts the vibration modes that fit into a cavity, and a quantum system with $d$ levels can oscillate at up to $d(d-1)/2$ frequencies without decay. For systems that jump at random between states, no such count was known: they oscillate only out of equilibrium, and the random timing of their jumps damps every oscillation. We find the missing count: if $Q$ is the ratio of frequency to decay rate, a system of $d$ states sustains fewer than $2(d-1)/(\pi Q)$ oscillations whose ratio exceeds $Q$. Nested cycles approach this bound, and measured oscillations reveal how large a hidden system must be.
\end{abstract}

\maketitle

Weyl's law counts the normal modes that fit into a cavity~\cite{weyl1912}; read in reverse, it lets one hear the area of a drum~\cite{kac1966}. The count is just as simple for springs and quantum systems: coupled springs or inductor--capacitor circuits hold one undamped oscillation for every two variables~\cite{goldstein2002,williamson1936}, and a quantum system with $d$ levels oscillates at up to $d(d-1)/2$ frequencies without decay~\cite{cohentannoudji1977}. For systems that jump at random between states, such as molecules that switch between shapes, no such count is known~\cite{dmitriev1946,karpelevich1951,barato2017}. These systems are described by the probabilities of their states, which obey linear equations like the positions of springs~\cite{schnakenberg1976}. But a spring oscillates by swinging through zero, and probabilities can never become negative. Probability can oscillate only by flowing around a cycle of states, like a hand around a dial; this flow requires a drive out of equilibrium~\cite{qian2000} and advances only by jumps. Copies that start together jump at random times and therefore drift out of step; in other words, every tick blurs the clock.

\begin{figure}[t]
\centering
\includegraphics[width=\columnwidth]{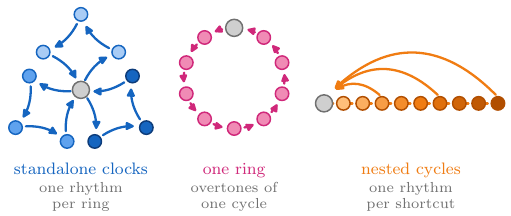}
\caption{\textbf{Three ways to hold many rhythms.} Left: standalone clocks, one-way rings that share one state, each with one sharp rhythm. Middle: one ring, whose rhythms are overtones, each less sharp than the one before. Right: nested cycles, one chain with shortcuts back to the start; each shortcut adds a rhythm. Every cycle passes through the gray state, and darker states are slower. At high sharpness, long chains of nested cycles need only half the states per rhythm of the other two, but they blur their rhythms together.}
\label{fig:ways}
\end{figure}

\begin{figure*}[t]
\centering
\includegraphics[width=\textwidth]{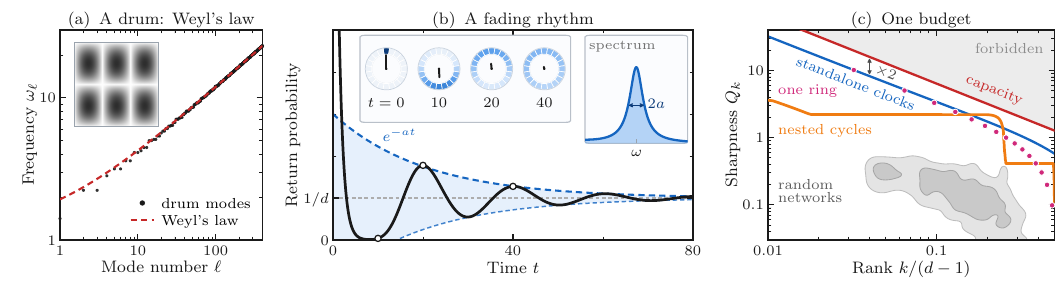}
\caption{\textbf{A drum's modes follow Weyl's law; a clock's rhythms fade and share one budget.} (a)~Frequency of the $\ell$th mode of a square drum (dots) and Weyl's law (red), whose height reveals the area of the drum; inset: one mode. (b)~Copies of a ring of 20 states, started together, drift out of step, so their chance of being back at the start oscillates and fades as $e^{-at}$. Left window: dials at four times, marked by circles; the clock hand, the average position of the copies, turns at the frequency $\omega$ and shrinks. Right window: the matching spectral line, with sharpness $Q=\omega/a$. (c)~Sharpness $Q_k$ of the $k$th sharpest rhythm against $k/(d-1)$: no network crosses the capacity (red), Eq.~\eqref{eq:law}, the counterpart of Weyl's law; standalone clocks and one ring stay about a factor of two below it, nested cycles come closest, and random networks lie far below (shading: 50\% and 90\% of their rhythms).}
\label{fig:budget}
\end{figure*}

Yet living cells keep time in exactly this way. Three proteins and adenosine triphosphate (ATP), the cell's energy carrier, suffice to sustain a circadian clock in a test tube~\cite{nakajima2005}, and molecular motors and enzymes step through cycles of chemical states~\cite{cao2015,liepelt2007,seifert2012,barato2016}. Physical reservoir computers exploit the same principle in hardware: they store the recent past in the fading oscillations of noisy devices~\cite{jaeger2004,maass2002,tanaka2019,hermans2010,sheldon2022,ehlers2025}. In each case, a single network must sustain several oscillations, which we call rhythms, at the same time; a motor couples a chemical cycle to a mechanical one~\cite{liepelt2007}, and a reservoir requires one rhythm for every frequency it remembers~\cite{hermans2010}~\SMsec{sec:reservoir}. Figure~\ref{fig:ways} sketches three ways to build such a network. Existing thermodynamic bounds constrain one rhythm at a time, or an average over all of them~\cite{kolchinsky2024,kolchinsky2026winding,barato2017,ohga2023,uhl2019,verbeken2026,xumeng2026,oberreiter2022,santolin2025,gu2026,kolchinsky2026elementary,shiraishi2026}. A recent ellipse theorem confines all eigenvalues of the matrix of jump rates to one ellipse, whose shape is set by the largest dissipation per jump on a single link~\cite{xu2025}. None of these bounds limits how many rhythms a network can sustain. How many rhythms can a stochastic network of $d$ states support, and at what cost?

We find the missing count. Let the sharpness $Q$ of a rhythm be the ratio of its frequency to its decay rate. That every tick blurs the clock suffices to give every network a rhythm capacity: a network of $d$ states supports $N$ rhythms sharper than $Q$ only if $NQ<2(d-1)/\pi$, whereas springs and quantum levels hold their oscillations at any sharpness. A ring of states devoted to a single rhythm, a standalone clock, needs about $\pi Q$ states~\cite{dmitriev1946,karpelevich1951}; in a shared network, each rhythm still costs at least half of that, and the costs of different rhythms add up. Networks of nested cycles approach this capacity at high sharpness, which yields a Weyl law for clocks, but only by blurring their rhythms together. Rhythms that remain distinguishable obey a square-root law instead: a band of frequencies holds only about $\sqrt d$ of them, whereas $d$ ordinary variables hold $d/2$.

Two consequences follow. First, each rhythm costs drive, the free energy supplied per turn of a cycle, nearly half of what a standalone clock requires, while the dissipated power can vanish~\cite{shiraishi2026}~\SMsec{sec:rate}. Second, like Weyl's law [Fig.~\ref{fig:budget}(a)], the capacity can be read in reverse: since a measurement can hide a rhythm but never create one~\SMsec{sec:inference}, a few measured rhythms reveal how many states a hidden machine must have and how strongly it must be driven. Together, these results do for stochastic clocks what Weyl's law did for waves: one count, valid for every network from a single enzyme to a reservoir computer, fixes how many rhythms fit and lets one hear the size of a clock.

\runin{Setup.}We consider a molecule that jumps at random between $d$ states with fixed rates and without memory: a Markov network, whose links are the possible jumps [Fig.~\ref{fig:ways}]. Its probabilities evolve as $\dot{\bm p}=W\bm p$, where $W_{ij}$ is the rate of jumps from $j$ to $i$ and every column of $W$ sums to zero. Every correlation function, which relates the state at one time to the state a time $\tau$ later, is then a sum of fading terms, one for each eigenvalue of $W$. We define a rhythm as a term that oscillates at frequency $\omega$ while it decays at rate $a$ [Fig.~\ref{fig:budget}(b)]. It appears as a spectral line at $\omega$ with half-width $a$ [Fig.~\ref{fig:budget}(b), right window], and as a complex eigenvalue pair $-a\pm i\omega$ of $W$. Its sharpness $Q=\omega/a$ measures how long it keeps time: the amplitude falls by a factor $e$ after $Q/2\pi$ periods. (Here, $Q$ is twice the optical $Q$ factor.) At equilibrium, every link carries equal flows in both directions, all eigenvalues of $W$ are real, and there are no rhythms at all~\cite{qian2000}. Rhythms therefore require a drive, the free energy supplied per turn of a cycle (its cycle affinity), in units of the thermal energy $k_BT$.

Even a single rhythm has a price. The sharpest rhythm that $d$ states can sustain belongs to a one-way ring with equal rates, for which $Q=\cot(\pi/d)\approx d/\pi$~\cite{dmitriev1946,karpelevich1951,runnenburg1962,kellogg1978}. One turn of such a ring consists of $d$ random waits in a row, whose total duration scatters by a fraction $1/\sqrt d$; the copies stay in step for a number of turns proportional to $d$. Faster jumps shorten the period and the lifetime by the same factor, so only additional states extend the number of coherent periods. A rhythm of sharpness $Q$ therefore requires about $\pi Q$ states and a drive above $2\pi Q\,k_BT$ per turn~\cite{barato2017,ohga2023}. For $Q=10$, this means at least 32 states and, at any size, at least $63\,k_BT$. We call the smallest such ring the standalone clock of one rhythm.

The origin of this price is that every tick blurs the clock. In the stationary state, probability flows across every link, generally in both directions. The total flow between states $i$ and $j$, the traffic $T_{ij}$ (or activity), smooths out differences between them. The net flow, the circulation $J_{ij}$ (or net current), carries probability around cycles and thereby advances the phase. Since both are built from the same nonnegative flows, the circulation can never exceed the traffic, $|J_{ij}|\le T_{ij}$. A clock therefore cannot turn without blurring, and all results below follow from this inequality.

\runin{Main result.}All rhythms of a network draw on a single budget that is set by its size. For any $k$ rhythms,
\begin{equation}
\sum_{j=1}^{k}\frac{1}{Q_j}\ \ge\ \frac{\pi k^{2}}{2(d-1)}.
\label{eq:capacity}
\end{equation}
In words, the relative widths $1/Q$ of any $k$ rhythms add up to at least an amount that grows like $k^2$. Consequently, a network supports $N$ rhythms sharper than $Q$ only if
\begin{equation}
NQ\ <\ \frac{2(d-1)}{\pi},
\label{eq:law}
\end{equation}
and its $k$th sharpest rhythm satisfies $Q\le2(d-1)/(\pi k)$ [Fig.~\ref{fig:budget}(c)]. Counted in periods, $n=Q/2\pi$, Eq.~\eqref{eq:law} reads $d-1>\pi^2Nn$: each rhythm requires nearly ten states for every period it lasts, compared with about twenty for standalone clocks.

The quadratic growth in $k$ is precisely what single-rhythm bounds miss. Such bounds would allow $N$ rhythms of sharpness $Q$ on about $\max(2N,\pi Q)$ states, two per rhythm or one long ring, whereas Eq.~\eqref{eq:law} requires $\pi NQ/2$. For one rhythm the single-rhythm bound is stronger, for two rhythms both are comparable, and from three comparably sharp rhythms onward the shared budget dominates. For example, five rhythms sharper than 10 require at least 80 states, although each of them alone fits in 32; five standalone clocks use 156.

\runin{Why the bound holds.}First, give each state any phase, like a position on a dial: the circulation turns this pattern, while the traffic smooths it. Writing the sharpness of each rhythm as a damping angle, $\arccot Q_j$, the angles of the $k$ sharpest rhythms add up to at least the angles $\alpha_j$ of the $k$ best ratios of turning to smoothing, $\cot\alpha_j$~\cite{horn1959,wang2020,furtado2001}. Second, these angles fix the roots $e^{-2i\alpha_j}$ of a polynomial whose coefficients are sums over the spanning trees of the network, the subnetworks that reach every state without closing a loop~\cite{chaiken1982}; none of its coefficients is negative. Third, such a polynomial cannot, on average, crowd its roots near the positive real axis~\cite{eremenko2016}, where sharp rhythms would require them. Counting the roots gives $\sum_{j\le k}\arccot Q_j\ge\pi k^2/[2(d-1)]$, which implies Eq.~\eqref{eq:capacity} because $\arccot Q<1/Q$; the Supplemental Material gives the full proof~\SMsec{sec:counting}. Since the argument uses only nonnegative flows, the capacity applies to any linear dynamics that keeps populations nonnegative~\cite{farina2000}, such as ion channels, enzymes, and compartment models, with decay measured from the slowest mode. For matrices with nonnegative entries, it is a new spectral condition, not implied by the classical ones~\cite{johnson1981,holtz2005}~\SMsec{sec:niep}.

\begin{figure}[t]
\centering
\includegraphics[width=\columnwidth]{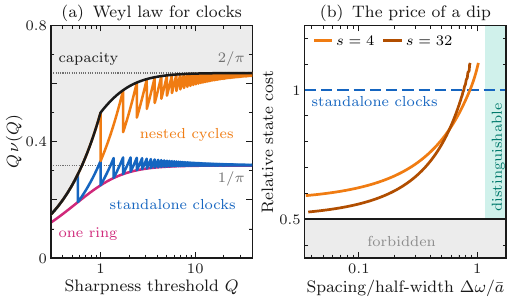}
\caption{\textbf{Sharing saves at most half of the states, and only with blurred rhythms.} (a)~How densely sharp rhythms can be packed: nested cycles reach the capacity at high sharpness, standalone clocks and one ring only half. (b)~The state cost of nested cycles relative to standalone clocks, as their rhythms move apart. Nested cycles lose their advantage before neighboring rhythms become distinguishable (shaded). Each curve is drawn until it clearly exceeds the cost of standalone clocks. Gray: forbidden.}
\label{fig:sharing}
\end{figure}

\runin{Sharing states.}Compared with one standalone clock per rhythm, sharing states can at most halve the price, and long chains of nested cycles indeed halve it at high sharpness. Standalone clocks, one-way rings that share a single state and run at very different speeds~\SMsec{sec:counting}, spend about $\pi Q$ states per rhythm, twice the capacity bound. A single large ring does no better, because its additional rhythms are overtones, each less sharp than the one before~\SMsec{sec:theta}. Along a chain of nested cycles [Fig.~\ref{fig:ways}], the jumps grow ever slower, and after every $s$ states a shortcut leads back to the start. Each shortcut closes a new cycle, and each cycle adds a rhythm of its own rather than an overtone, since short and long excursions share the start of their route but return at different times. Deep in a long chain, every new rhythm is a slowed copy of the previous one and is therefore equally sharp [Fig.~\ref{fig:budget}(c)]. As a result, the largest number of rhythms sharper than $Q$ per state in large networks, $\nu(Q)$, obeys
\begin{equation}
\lim_{Q\to\infty}Q\,\nu(Q)\ =\ \frac{2}{\pi},
\label{eq:weyl}
\end{equation}
a Weyl law for clocks, twice the $1/\pi$ of standalone clocks [Fig.~\ref{fig:sharing}(a)]: like Weyl's law for modes, it fixes the number of sharp rhythms per state. Below $Q=1$, nested cycles even reach the absolute maximum of one rhythm per two states. The gain appears only in very long chains, however, because the blocks of $s$ states near either end of a chain contribute few sharp rhythms~\SMsec{sec:finitechains}.

\runin{Distinguishable rhythms.}Nested cycles pay for this gain with blur. We call two rhythms distinguishable when their spacing exceeds $(a_1+a_2)/\sqrt3$, the Sparrow criterion for a dip between two equal lines~\cite{sparrow1916}. The rhythms of nested cycles lie far closer together than their linewidths, and the chain loses its advantage over standalone clocks before any dip appears [Fig.~\ref{fig:sharing}(b)]. To test whether other networks do better, we optimized all rates of networks of 21, 31, and 41 states in 720 runs. The optimized networks hold one rhythm more than standalone clocks when rhythms may blur, but not a single additional rhythm once every rhythm must be distinguishable~\SMsec{sec:optimized}. The reason lies at branch points: a branch whose rhythm is distinguishable returns out of phase with the others and acts as a leak, whereas branches with blurred rhythms recombine coherently. We therefore conjecture that distinguishable rhythms are additive, each costing a standalone clock up to rounding, as they provably are on widely separated time scales~\SMsec{sec:additive}.

Distinguishable rhythms obey a square-root law. $N$ of them whose frequencies span a ratio $R$ require
\begin{equation}
d-1\ >\ \frac{\pi\,(N-1)^{2}}{\sqrt{3}\,\ln R}
\label{eq:sqrt}
\end{equation}
states, and rings reach this scaling~\SMsec{sec:band}. A fixed band therefore holds only about $\sqrt d$ distinguishable rhythms of $d$ probabilities, whereas $d$ ordinary variables, such as the positions and velocities of springs, hold $d/2$~\cite{williamson1936}. The nonnegativity of probabilities thus costs a square root.

\runin{Energy cost.}Rhythms are also paid for in drive. If no jump dissipates more than $\varepsilon\,k_BT$, every circulation is at most a fraction $\mu=\tanh(\varepsilon/2)$ of its traffic; the same factor caps every frequency in the ellipse theorem~\cite{xu2025}. Turning is then weak compared with blurring, and $N$ rhythms require $d-1\gtrsim\pi NQ/(2\mu)$ states. One ATP, about $20\,k_BT$~\cite{milo2016}, spread over 20 jumps gives $\varepsilon\approx1$ and thus more than doubles the state cost~\SMsec{sec:weakstep}.

\begin{figure}[t]
\centering
\includegraphics[width=\columnwidth]{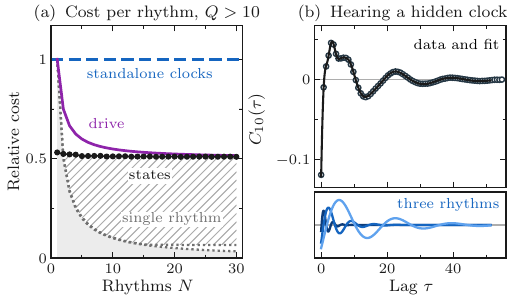}
\caption{\textbf{Rhythms cost states and drive, and measured rhythms reveal both.} (a)~States and drive per rhythm, relative to standalone clocks: when many sharp rhythms share a network, each still costs at least about half a standalone clock, whereas single-rhythm bounds (dotted) would allow a few percent of its states and almost none of its drive. Gray: forbidden; hatched: forbidden only by the shared budget. (b)~A simulated hidden machine seen through a coarse readout: three rhythms (below) in its measured correlation (circles, with fit) reveal a minimum size and drive~\SMsec{sec:inference}.}
\label{fig:costs}
\end{figure}

Moreover, every rhythm pays its own drive. Each driven cycle offers a ladder of overtones whose sharpness its drive caps, and $N$ rhythms must be paid from the $N$ best rungs of all ladders together. Let the total drive $\Ft$ measure the free energy that drives the whole network; it never exceeds the summed drives of its cycles~\cite{schnakenberg1976,kalpazidou2006,ahuja1993,SM}. Then $N$ rhythms sharper than $Q>1$ require
\begin{equation}
\Ft\ >\ \pi(N+1)\left(Q-\frac{1}{Q}\right).
\label{eq:drive}
\end{equation}
For one rhythm, Eq.~\eqref{eq:drive} nearly recovers the price of a single ring~\cite{ohga2023}. For many rhythms, the price per rhythm approaches $\pi Q=2\pi^2n$: one ATP, about $20\,k_BT$, for each period it lasts, half of what a standalone clock requires [Fig.~\ref{fig:costs}(a)]. Five rhythms sharper than 10, for example, require close to $190\,k_BT$, whereas one requires 63. If each driven cycle burns one ATP per turn, the network therefore needs at least ten driven cycles. In the nested cycles we examined, sharing did not save drive~\SMsec{sec:affcost}.

In contrast, power is free, because a rhythm requires free energy per turn rather than per unit time. To see this, we let the molecule park by adding a parking state that every ring state enters at a small fixed rate and that refills the rings only rarely. The molecule then spends almost all of its time parked, and the dissipated power vanishes. Nevertheless, every rhythm keeps its frequency exactly, and its decay rate grows only by the small rate of parking. It appears in full whenever a prepared molecule relaxes, while its weight in stationary correlations vanishes with the power~\SMsec{sec:rate}. Consequently, no law can bound the sharpness of rhythms by the dissipated power. Reference~\cite{shiraishi2026} gave such a counterexample for one rhythm, against a conjecture of Ref.~\cite{oberreiter2022}; parking, which localizes the molecule as in Ref.~\cite{gu2026}, extends it to all rhythms at once.

\runin{Hearing a hidden clock.}A measurement can hide a rhythm but never create one. Every fading oscillation in a measured correlation function, or in the response to a weak perturbation, is a rhythm of the hidden network, however coarse the readout~\SMsec{sec:inference}. Measured rhythms therefore obey the capacity: solved for $d$, Eq.~\eqref{eq:capacity} shows that measured sharpnesses $Q_1\ge\cdots\ge Q_k$ certify
\begin{equation}
d\ \ge\ 1+\left\lceil\frac{\pi m^{2}}{2\sum_{j\le m}1/Q_j}\right\rceil\qquad(\text{any } m\le k)
\label{eq:hearing}
\end{equation}
hidden states, along with $d\ge2k+1$ and $d>\pi Q_1$. The drive law likewise certifies a minimum total drive~\SMsec{sec:inference}. No model of the machine enters, and the spectral lines need not be resolved. Equation~\eqref{eq:hearing} thus bounds the number of hidden states from data alone, a dimension witness for classical stochastic dynamics~\cite{brunner2008,wolf2009}, and it extends thermodynamic inference~\cite{seifert2019,horowitz2020} to many rhythms. Kac asked whether one can hear the shape of a drum~\cite{kac1966}; one can hear how large a clock must be.

Several rhythms reveal far more than one. We simulated a hidden machine of 42 states with three driven cycles and a total drive of $126\,k_BT$. We observed only which third of its cycle it occupies, much like a motor whose position is tracked but not its chemistry [Fig.~\ref{fig:costs}(b)]. A standard fit of fading oscillations~\cite{hua1990}, with confidence limits from resampling whole trajectories~\cite{efron1979}, identifies three rhythms. At 95\% confidence, they certify at least 17 states and $39.8\,k_BT$ of drive, whereas the sharpest rhythm alone certifies only 13 states and $24.1\,k_BT$. In 600 simulated data sets, the certified number of states never exceeded the true one~\SMsec{sec:protocol}; this guarantee rests on lower confidence limits for the sharpnesses rather than on point estimates. The same analysis applies to single-molecule traces of motors, enzymes, and ion channels, and to stochastic reservoir computers~\cite{ehlers2025}: a reservoir that holds $N$ sharp memory rhythms requires the states and the drive of Eqs.~\eqref{eq:law} and \eqref{eq:drive}~\SMsec{sec:reservoir}.

\runin{Conclusion.}Weyl's law counts the modes of a cavity; we have found the corresponding count for clocks that run on random jumps. Because every jump that advances such a clock also blurs it, all its rhythms draw on one budget: a network of $d$ states holds fewer than $2(d-1)/(\pi Q)$ rhythms sharper than $Q$, whereas springs and quantum levels hold theirs at any sharpness. Sharp rhythms therefore compete for states. Compared with one standalone clock per rhythm, sharing states saves at most half the states, and the only known way to do so blurs the rhythms together; rhythms that stay distinguishable fit only about $\sqrt d$ into a band of frequencies. Rhythms also need drive, free energy per turn of a cycle, but not power: they survive even when the dissipated power vanishes. The shape of a drum cannot always be heard~\cite{gordon1992}; the size of a clock can, at least from below: since a measurement can hide a rhythm but never create one, a few rhythms in any measured signal reveal how many states a hidden clock must have and how strongly it must be driven. Open questions remain, in particular whether distinguishable rhythms are additive, whether the drive bound can be reached, and whether open quantum systems have a capacity of their own.

\runin{Acknowledgments.}Claude Opus 5.5 (Anthropic) assisted with manuscript preparation, checks of derivations and references, and simulation and plotting code. Markus Baumann reviewed the outputs and takes responsibility for the final content.

\runin{Data availability.}The bounds are analytical; all code and data, including exact rates of the certified networks and a verification script, are provided with the Supplemental Material~\SM.

\let\SavedAddContentsLine\addcontentsline\renewcommand{\addcontentsline}[3]{}
\bibliography{refs}
\let\addcontentsline\SavedAddContentsLine

\clearpage
\onecolumngrid
\setcounter{page}{1}
\renewcommand{\thepage}{S\arabic{page}}
\setcounter{equation}{0}
\renewcommand{\theequation}{S\arabic{equation}}
\setcounter{section}{0}
\setcounter{secnumdepth}{3}
\setcounter{theorem}{0}
\setcounter{lemma}{0}
\setcounter{corollary}{0}

\begin{center}
{\large\bfseries Supplemental Material:\\[1pt] Can One Hear the Size of a Clock?}\\[10pt]
Markus Baumann\,\orcidlink{0009-0007-3575-1006} and Claudia Linnhoff-Popien\,\orcidlink{0000-0001-6284-9286}\\[3pt]
{\small\itshape Quantum Applications and Research Laboratory (QAR-Lab),\\ Ludwig-Maximilians-Universit\"at M\"unchen, Oettingenstra{\ss}e 67, 80538 M\"unchen, Germany}
\end{center}

\medskip
This Supplemental Material gives the proofs and computations behind the Letter. Each section starts with its result. Equations, theorems, and lemmas numbered S1, S2, \dots\ belong to this Supplemental Material; plain numbers refer to the Letter.

\paragraph*{Notation.} $W$ is the matrix of jump rates: $W_{ij}\ge0$ ($i\ne j$) is the rate of the jump $j\to i$, and every column of $W$ sums to zero. A rhythm is a complex eigenvalue pair $-a\pm i\omega$ of $W$ with $a,\omega>0$, counted once and with multiplicity. Its sharpness is $Q=\omega/a$ and its damping angle is $\delta=\arctan(a/\omega)=\arccot Q$: sharp rhythms have small angles. Rhythms are numbered by sharpness, $Q_1\ge Q_2\ge\cdots$, so that $\delta_1\le\delta_2\le\cdots$; a threshold sharpness is written $Q_0$ (simply $Q$ in the Letter). Other symbols are defined where they first appear; symbols introduced inside a single proof or construction are local to it.

\tableofcontents

\section{The rhythm capacity}\label{sec:counting}

\begin{theorem}[Rhythm capacity]\label{thm:counting}
For every network of $d\ge2$ states and every $k$ up to its number of rhythms,
\begin{equation}
\sum_{j=1}^{k}\delta_j\ \ge\ \frac{\pi k^2}{2(d-1)}.
\label{eq:S-counting}
\end{equation}
\end{theorem}

\noindent This is the sharp form quoted in the Letter. Any other choice $I$ of $k$ rhythms has at least the same angle sum, and $\delta\le1/Q$, so
\begin{equation}
\sum_{j\in I}\frac{1}{Q_j}\ \ge\ \sum_{j\in I}\delta_j\ \ge\ \frac{\pi k^2}{2(d-1)},
\label{eq:S-reldamp}
\end{equation}
which is Eq.~\eqref{eq:capacity} of the Letter.

\begin{corollary}[Counting rhythms]\label{cor:count}
The $k$th sharpest rhythm has $Q_k\le\cot\{\pi k/[2(d-1)]\}$, and $N$ rhythms sharper than $Q_0$ need
\begin{equation}
d-1\ >\ \frac{\pi N}{2\arccot Q_0}\ >\ \frac{\pi NQ_0}{2}.
\label{eq:S-cost}
\end{equation}
\end{corollary}

\begin{proof}
The angles increase, so $k\delta_k\ge\sum_{j\le k}\delta_j\ge\pi k^2/[2(d-1)]$. If $N$ rhythms are sharper than $Q_0$, then $\pi N/[2(d-1)]\le\delta_N<\arccot Q_0<1/Q_0$.
\end{proof}

\subsection{The benchmark: standalone clocks}

One rhythm sharper than $Q_0$ needs a ring of $n_0$ states, the smallest $n_0$ with $\cot(\pi/n_0)>Q_0$~\cite{dmitriev1946,karpelevich1951,runnenburg1962,kellogg1978}; then $n_0-1\le\pi/\arccot Q_0<n_0$. Standalone clocks hold $N$ such rhythms on $N(n_0-1)+1$ states: $N$ one-way rings of $n_0$ states that share one state. Label the rings from slowest ($c=1$) to fastest ($c=N$). The $n_0-1$ own states of ring $c$ are left at rate $v_c$, with $v_1\ll\cdots\ll v_N$, and the shared state enters ring $c$ at rate
\begin{equation}
w_N=v_N,\qquad w_c=v_c\Big[1+(n_0-1)\sum_{c'>c}\frac{w_{c'}}{v_{c'}}\Big].
\label{eq:S-entry}
\end{equation}
These graded rates, $w_c=v_cn_0^{N-c}$ in closed form, make the mean time between leaving ring $c$ and entering it again, including the waits in the shared state and the excursions through faster rings, equal to $1/v_c$ as long as no slower ring is entered. Entries into slower rings act as leaks, with probability about $\sum_{c'<c}(v_{c'}/v_c)\,n_0^{c-c'}$ per return, which vanishes once the speed ratios outgrow these powers of $n_0$. This nearly exponential delay acts as the missing $n_0$th stage of ring $c$, so every ring tends to $Q=\cot(\pi/n_0)$ as the speed ratios grow.

Equal entry rates, $w_c=v_c$, do not work: every return to the shared state then starts a run of excursions through the faster rings, which delays ring $c$ by about one period per faster ring. For three rings of 8 states, equal rates give damping angles $0.74$, $0.61$, and $0.39$ (slowest to fastest) at every speed ratio from $10^2$ to $10^8$, against $\pi/8\approx0.393$. Graded rates give $0.39273$, $0.39285$, and $0.39282$ at ratio $10^4$, and $\pi/8$ to within $2\times10^{-8}$ at $10^8$ (\texttt{repro/standalone\_clocks.py}).

\paragraph*{Sharing saves at most a factor of two.} By Eq.~\eqref{eq:S-cost}, every network needs $d-1>\pi N/(2\arccot Q_0)\ge N(n_0-1)/2$: at least half of the unshared states of standalone clocks. For one rhythm the ring bound $d\ge n_0$ is stronger, for two Eq.~\eqref{eq:S-cost} gives nearly the same count, and from three rhythms on the shared budget dominates. Seven rhythms just above $Q_0=10$, for example, need $d-1>110.3$ states, so they do not fit on 100 states, although each alone fits on 32.

\subsection{Proof of Theorem~\ref{thm:counting}}

The proof has four steps: compare turning with smoothing, turn the ratios into a bound on the damping angles, encode the ratios in a polynomial with nonnegative coefficients, and count its roots near the positive axis.

\paragraph*{Step 1: turning and smoothing.} Let $W$ be irreducible with stationary distribution $\bm\pi>0$ (components $\pi_i$), and put $\Pi=\mathrm{diag}(\bm\pi)$ and $\Phi=W\Pi$. The entry $\Phi_{ij}=W_{ij}\pi_j$ is the stationary flow from $j$ to $i$, and $\Phi$ has zero row and column sums. On a link $e=\{i,j\}$ the traffic is $T_e=\Phi_{ij}+\Phi_{ji}$ and the circulation is $J_e=\Phi_{ij}-\Phi_{ji}$. The Hermitian matrices $L=-(\Phi+\Phi^{\mathsf T})/2$ and $K=(\Phi-\Phi^{\mathsf T})/(2i)$ act on a pattern of phases $y_i=|y_i|e^{i\psi_i}$ as
\begin{equation}
y^*Ly=\frac12\sum_eT_e|y_i-y_j|^2,\qquad y^*Ky=\sum_eJ_e|y_i||y_j|\sin(\psi_j-\psi_i).
\label{eq:S-forms}
\end{equation}
The traffic smooths differences; the circulation turns phases. On the $(d-1)$-dimensional space orthogonal to constants, $L$ is positive definite, so the ratios of turning to smoothing, the generalized eigenvalues of $Ky=\rho Ly$, are real. Write them as $\rho_j=\cot\alpha_j$ with $0<\alpha_1\le\cdots\le\alpha_{d-1}<\pi$. Because $L$ is real and $\bar K=-K$, they come in opposite pairs. The $\alpha_j$ are auxiliary angles, not the damping angles of individual rhythms.

\paragraph*{Step 2: from ratios to damping angles.}

\begin{lemma}[Eigenvalue arguments~\cite{horn1959,wang2020,furtado2001}]\label{lem:args}
Let $X=A+iB$ with $A,B$ Hermitian and $A\succ0$. Let $\tilde\rho_1\ge\cdots\ge\tilde\rho_m$ be the eigenvalues of $A^{-1/2}BA^{-1/2}$ and $\theta_1\ge\cdots\ge\theta_m$ the arguments of the eigenvalues of $X$, taken in $(-\pi/2,\pi/2)$. Then $\sum_{j\le p}\theta_j\le\sum_{j\le p}\arctan\tilde\rho_j$ for $1\le p\le m$, with equality for $p=m$.
\end{lemma}

\begin{proof}
The identity $\det X=\det A\prod_j(1+i\tilde\rho_j)$ gives the full sum; along $A+i\beta B$, $0\le\beta\le1$, all eigenvalues stay in the right half-plane, so no multiple of $2\pi$ appears. For $p<m$, restrict $X$ to a Schur invariant subspace of its $p$ largest arguments. The restriction keeps a positive definite Hermitian part, and by the min--max principle its ratios are at most $\tilde\rho_1,\dots,\tilde\rho_p$.
\end{proof}

\noindent Apply the lemma to $X=-\Pi^{-1/2}W\Pi^{1/2}$ on the complement of $\bm\pi^{1/2}$, which carries all nonzero eigenvalues of $-W$. With $y=\Pi^{-1/2}x$, its parts $A$ and $B$ give $x^*Ax=y^*Ly$ and $x^*Bx=-y^*Ky$; the sign does not matter because the ratios come in opposite pairs. Each rhythm contributes the argument $\arctan Q_j=\pi/2-\delta_j$, and each ratio the argument $\arctan\rho_j=\pi/2-\alpha_j$. Hence
\begin{equation}
\sum_{j=1}^{k}\delta_j\ \ge\ \sum_{j=1}^{k}\alpha_j .
\label{eq:S-aux}
\end{equation}
The bound holds for partial sums; no rhythm is matched to a single ratio.

\paragraph*{Step 3: a polynomial with nonnegative coefficients.} Let $p(z)$ be a principal cofactor of $-(z\Phi+\Phi^{\mathsf T})$. By the directed matrix-tree theorem~\cite{chaiken1982} it is a sum over spanning trees of products of factors $z\Phi_{ij}+\Phi_{ji}$, so no coefficient is negative, and the constant and leading coefficients are positive. Since $-(z\Phi+\Phi^{\mathsf T})=(1+z)L+i(1-z)K$, the determinant on the nonconstant subspace gives, with $C>0$,
\begin{equation}
p(z)=C\prod_{j=1}^{d-1}\bigl[(1+i\rho_j)+z(1-i\rho_j)\bigr],
\label{eq:S-kirchhoff}
\end{equation}
whose roots $z_j=e^{-2i\alpha_j}$ lie on the unit circle. A small angle $\alpha_j$ puts a conjugate pair of roots near the positive real axis.

\paragraph*{Step 4: the roots avoid the positive axis on average.}

\begin{lemma}[Averaged root count~\cite{eremenko2016}]\label{lem:ef}
For a polynomial of degree $n$ with nonnegative coefficients and positive constant and leading coefficients,
\begin{equation}
\sum_{p(z)=0}\max\{\varphi-|\arg z|,0\}\ \le\ \frac{n\varphi^2}{2\pi},\qquad 0<\varphi\le\pi,
\label{eq:S-ef}
\end{equation}
with equality for $z^n+1$ at $\varphi=2\pi k/n$.
\end{lemma}

\begin{proof}
Let $\mathcal N(\vartheta)$ count the roots with $|\arg z|<\vartheta$. The argument principle gives $\mathcal N(\vartheta)=[n\vartheta-\Delta(\vartheta)]/\pi$, where $\Delta(\vartheta)$ is the change of $\arg p(re^{i\vartheta})$ as $r$ runs from zero to infinity. Integrating the Cauchy--Riemann equations in $u=\ln r$ gives $\int_0^{\varphi}\Delta\,d\vartheta=\int_{-\infty}^{\infty}[\ln p(e^u)-\ln|p(e^{u+i\varphi})|]\,du\ge0$, because nonnegative coefficients give $|p(re^{i\vartheta})|\le p(r)$. Integrating $\mathcal N(\vartheta)$ over $\vartheta$ from zero to $\varphi$ proves Eq.~\eqref{eq:S-ef}.
\end{proof}

\noindent Obrechkoff's pointwise count, $\mathcal N(\vartheta)\le2n\vartheta/\pi$, integrated over $\vartheta$, would give twice the right side of Eq.~\eqref{eq:S-ef} and only half of Theorem~\ref{thm:counting}; the averaging is what makes the sharp constant possible.

\paragraph*{Assembly.} A network with $k$ rhythms has $2k\le d-1$. The $k$ smallest angles $\alpha_j\le\pi/2$ select $k$ conjugate root pairs, each contributing $4\max\{x-\alpha_j,0\}$ to Eq.~\eqref{eq:S-ef} with $n=d-1$ and $\varphi=2x\le\pi$. Therefore
\begin{equation}
\sum_{j=1}^{k}\alpha_j\ \ge\ kx-\frac{(d-1)x^2}{2\pi},
\label{eq:S-assembly}
\end{equation}
which is largest at $x=\pi k/(d-1)$, where it equals $\pi k^2/[2(d-1)]$. With Eq.~\eqref{eq:S-aux} this proves Theorem~\ref{thm:counting} for irreducible networks. A reducible network becomes irreducible when arbitrarily small rates are added; its nonreal eigenvalues move continuously, so the bound holds in the limit, one-way jumps included.

\subsection{Nonnegative matrices}\label{sec:niep}

A real linear system $\dot x=Mx$ keeps populations nonnegative exactly when $M_{ij}\ge0$ for $i\ne j$~\cite{berman1994,farina2000}. For irreducible $M$ with dominant real eigenvalue $\mu_1$ and positive left eigenvector $\bm u$, the matrix $W=D(M-\mu_1I)D^{-1}$ with $D=\mathrm{diag}(\bm u)$ is a matrix of jump rates with the shifted spectrum. Theorem~\ref{thm:counting} therefore holds for $M$, with decay rates measured relative to $\mu_1$; reducible systems follow by continuity. The assumption concerns the full linear dynamics: it does not extend the result to nonlinear models or to ordinary (signed) variables.

Every spectrum of an $n\times n$ nonnegative matrix is realized by one with constant row sums equal to its Perron root $r$~\cite{johnson1981}, whose transpose minus $rI$ is a matrix of jump rates. Its nonreal eigenvalues $\mu_j$ with $\mathrm{Im}\,\mu_j>0$ therefore obey, for the $k$ smallest angles,
\begin{equation}
\sum_{j\le k}\arctan\frac{r-\mathrm{Re}\,\mu_j}{\mathrm{Im}\,\mu_j}\ \ge\ \frac{\pi k^2}{2(n-1)}.
\label{eq:S-niep}
\end{equation}
The classical conditions do not imply this one. The list $\sigma_W=\{0,\,-1\pm2.3i,\,-(1\pm2.3i)/400,\,-0.0055,\,-2.5,\,-150\}$ violates Eq.~\eqref{eq:S-niep} at $n=8$, $k=2$, with angle sum $2\arctan(1/2.3)\approx0.820<2\pi/7\approx0.898$. Yet $-\sigma_W+r$ satisfies Newton's inequalities~\cite{holtz2005} for every $r\ge0$ (in exact arithmetic), $\sigma_W+r$ satisfies the Karpelevi\v{c} conditions~\cite{dmitriev1946,karpelevich1951} for every $r\ge154.51$, and the trace inequalities~\cite{johnson1981,loewy1978} of all orders up to $10^5$ hold at the shifts tested. Each pair passes the single-rhythm test; together they fail.

\section{Nested cycles and the Weyl law for clocks}\label{sec:theta}

Let $N^*(d,Q_0)$ be the largest number of rhythms sharper than $Q_0$ in a network of $d$ states, and $\nu(Q_0)=\lim_{d\to\infty}N^*(d,Q_0)/(d-1)$ (everything below holds for the lower and upper limits alike). Corollary~\ref{cor:count} gives $\nu(Q_0)\le(2/\pi)\arccot Q_0$, and trivially $\nu\le1/2$. This section shows that chains of nested cycles reach the first bound at high sharpness:
\begin{equation}
\lim_{Q_0\to\infty}Q_0\,\nu(Q_0)=\frac{2}{\pi},
\label{eq:S-sharp}
\end{equation}
which is Eq.~\eqref{eq:weyl} of the Letter.

\subsection{The chain and its rhythms}

\paragraph*{Construction.} Choose a block size $s\ge2$, a number of blocks $B$, $h>0$, and $0<\kappa<1$, and put $q=e^{-h}$, $H=sh$, and $p=1-e^{-\kappa H}$. The chain $\mathcal T(s,h,\kappa,B)$ has $d=sB$ states $j=0,\dots,d-1$; state $j=0$ is the start. State $j$ is left at rate $q^j$. Inside a block the jump goes to $j+1$. From the last state of a block the chain returns to the start with probability $p$ and continues into the next block otherwise; from its last state it always returns. Shifting the chain by one block multiplies all rates by $e^{-H}$.

\paragraph*{Why it works.} Excursions of different lengths share the start of their route but return at different times, so each block closes a new cycle and adds a rhythm of its own, whereas a ring adds only overtones [Figs.~\ref{fig:ways} and \ref{fig:budget}(c)]. By self-similarity, every new rhythm is a slowed copy of the previous one.

\begin{theorem}[One rhythm per interior block]\label{thm:theta}
Fix $s\ge2$, $0<\kappa<1$, and $0<\eta\le\pi/(4s)$. For every sufficiently small $h>0$ there is a constant $C$, independent of $B$, such that the chain has at least $B-C$ rhythms with damping angles in $(\delta_0-\eta,\delta_0+\eta)$, where $\delta_0=\pi/(2s)$.
\end{theorem}

\paragraph*{Proof of the Weyl law.} For large $Q_0$, take the smallest $s\ge2$ with $\cot[\pi/(2s)]>Q_0$, choose $\eta$ with $\delta_0+\eta<\arccot Q_0$, and fix $h$ as in the theorem. All $B-C$ rhythms are then sharper than $Q_0$, so $\liminf_{B\to\infty}N^*(sB,Q_0)/(sB-1)\ge1/s$; weakly linked extra states fill in the other values of $d$. Since $s\sim\pi Q_0/2$, this and Corollary~\ref{cor:count} give Eq.~\eqref{eq:S-sharp}. For $Q_0<1$, $s=2$ gives angles near $\pi/4<\arccot Q_0$ and density $1/2$, the trivial maximum. Standalone clocks reach only $Q_0\nu\to1/\pi$ [Fig.~\ref{fig:sharing}(a)], and so does one uniform ring of $d$ states: its overtones have $Q_m=\cot(\pi m/d)$, which exceeds $Q_0$ only for $m<d\arccot(Q_0)/\pi$.

\paragraph*{The rhythms are blurred.} Neighboring rhythms sit in adjacent cells of width $H$ in $\ln|\lambda|$ (Sec.~\ref{sec:thetaproof}), so their frequencies differ by $|\ln(\omega_2/\omega_1)|<2H+\ln[\cos(\delta_0-\eta)/\cos(\delta_0+\eta)]$, which vanishes as $\eta,h\to0$, while each relative width stays above $\tan(\delta_0-\eta)$. Neighbors are thus spaced by about $sh\,\omega$, far less than their linewidth. In the 300-state chain $\mathcal T(3,0.04,1/2,100)$, 19 rhythms fall into one decade of frequency, and at most three of them are pairwise distinguishable (Sec.~\ref{sec:resolved}). Equation~\eqref{eq:S-sharp} counts sharp rhythms, not distinguishable ones.

\subsection{Proof of Theorem~\ref{thm:theta}}\label{sec:thetaproof}

The proof has five steps: (i) reduce the eigenvalues to the return paths; (ii) compare the finite chain with a self-similar function $G$; (iii) expand $G$ in a Fourier series; (iv) show that two terms of this series dominate, which gives one zero of $G$ per logarithmic cell; (v) carry these zeros over to the finite chain, away from its two ends.

\paragraph*{(i) Return paths.} An excursion returns after $\ell+1$ blocks with probability $w_\ell=p(1-p)^\ell$ for $\ell<B-1$ and $w_{B-1}=(1-p)^{B-1}$. Every cycle passes through the start, so the expansion of $\det(\lambda I-W)$ has one term per return path, and
\begin{equation}
\det(\lambda I-W)=\Lambda(\lambda)\prod_{j=0}^{d-1}(\lambda+q^j),\qquad
\Lambda(\lambda)=1-\sum_{\ell=0}^{B-1}w_\ell\prod_{j=0}^{s(\ell+1)-1}\bigl(1+\lambda q^{-j}\bigr)^{-1}.
\label{eq:S-renewal}
\end{equation}
The possible poles of $\Lambda$ are real and negative, so its nonreal zeros are exactly the nonreal eigenvalues of $W$.

\paragraph*{(ii) Comparison function.} Write $P_n(y)=\prod_{r=0}^{n-1}(1+yq^r)$, $P(y)=P_\infty(y)$, and $f(y)=1-1/P(y)$. With $y_\ell=q\lambda e^{H(\ell+1)}$, the excursion through $\ell+1$ blocks contributes $1-1/P_{s(\ell+1)}(y_\ell)$ to $\Lambda$. Extending the blocks in both directions gives
\begin{equation*}
G(\lambda)=\sum_{\ell\in\mathbb Z}p\,e^{-\kappa H\ell}f(y_\ell),
\end{equation*}
which is exactly self-similar. Two elementary estimates control the products: for $0<h\le1/10$ and $|\arg y|\le\theta_{\max}<\pi$, with $\theta_+=\max(\theta_{\max}-\pi/2,0)$,
\begin{equation}
|P_n(y)|\ \ge\ \exp\!\Big[-\frac{2\ln2+2|\ln\cos\theta_+|}{h}\Big]\quad(n\le\infty),\qquad
\max\bigl\{|P(y)-1|,|1/P(y)-1|\bigr\}\le\frac{11|y|}{h}\quad(|y|\le h/4).
\label{eq:S-prod}
\end{equation}
The first holds because at most $2/h$ factors have $1/2<|y|q^r<2$ and the small factors form a geometric series; the second follows from $|\ln P(y)|\le2|y|/(1-q)\le1$.

\paragraph*{(iii) Fourier series.} Partial fractions of $1/P$ and Euler's product identity give, for $-1<\mathrm{Re}\,z<0$, the Mellin transform $\int_0^\infty y^{z-1}f(y)\,dy=-[\pi/\sin\pi z]\,P(-q^{1-z})/P(-q)$. For $\psi(u)=e^{-\kappa u}f(qe^u)$ this becomes the Fourier transform
\begin{equation}
\hat\psi(\xi)=q^{\kappa+i\xi}\,\frac{\pi}{\sin\pi(\kappa+i\xi)}\,\frac{P(-q^{1+\kappa+i\xi})}{P(-q)},
\label{eq:S-fourier}
\end{equation}
and Poisson summation gives, with $\lambda=e^x$,
\begin{equation}
G(e^x)=\frac{p}{H}\,e^{\kappa(x+H)}\sum_{j\in\mathbb Z}c_jZ^j,\qquad c_j=\hat\psi(-2\pi j/H),\qquad Z=e^{-2\pi ix/H}.
\label{eq:S-poisson}
\end{equation}
For an eigenvalue with damping angle $\delta$ in the upper half-plane, $\arg\lambda=\pi/2+\delta$ and $|Z|=\exp[(2\pi/H)(\pi/2+\delta)]$: the damping angle becomes a radius, and one logarithmic cell, an interval of length $H$ in $\ln|\lambda|$, becomes an annulus.

\paragraph*{(iv) Two terms dominate.} With $r$ the remainder of $j$ modulo $s$, put $E_j(\delta)=-2\pi^2|j|+\pi^2j+2\pi j\delta+\pi^2r(s-r)/s$. For small $h$,
\begin{equation*}
\frac{|c_jZ^j|}{c_0}\le4e^{E_j(\delta)/H}\quad(j\ne0),\qquad\frac{|c_jZ^j|}{c_0}\ge\sin(\pi\kappa)\,h^2e^{-4}\,e^{E_j(\delta)/H}\quad(j\ge1).
\end{equation*}
To see this, divide Eq.~\eqref{eq:S-fourier} by its value at zero. The sine factor lies between $\sin(\pi\kappa)e^{-2\pi^2|j|/H}$ and $4e^{-2\pi^2|j|/H}$, the radius gives $e^{2\pi j(\pi/2+\delta)/H}$, and the product factor lies between $h^2e^{-4}$ and $1$ times $\exp[\pi^2r(s-r)/(sH)]$. Its logarithm is $\sum_{n\ge0}g(h(n+1+\kappa))$ with $g(u)=\ln[|1-e^{2\pi ij/s}e^{-u}|/(1-e^{-u})]$, which is positive and decreasing, and $\int_0^\infty g\,du=\pi^2r(s-r)/s^2$. The sum is therefore at most $1/h$ times this integral and at least $1/h$ times its part above $h(1+\kappa)$; since $g(u)\le\ln(4/u)$ there, the missing part costs at most $(1+\kappa)\ln[4e/((1+\kappa)h)]$, so the factor stays above $[(1+\kappa)h/(4e)]^{1+\kappa}\ge h^2e^{-4}$.

For $0\le j\le s$ the exponent is $E_j=2\pi j\delta-\pi^2j^2/s$, so $E_0=0$ and $E_1=2\pi(\delta-\delta_0)$: these two terms exchange dominance at $\delta=\delta_0$. All other exponents are negative when $|\delta-\delta_0|\le\eta$: $E_j\le-(\pi^2/s)j(j-3/2)$ for $2\le j\le s$, a shift of $j$ by $s$ lowers $E_j$ by at least $\pi^2(s-3/2)$, and $E_j\le-2\pi^2|j|$ for $j<0$. Summing the geometric tails, for $h$ small enough (depending on $s,\kappa,\eta$, not on $B$),
\begin{equation}
\sum_{j\notin\{0,1\}}|c_jZ^j|\le\tfrac18c_0\ \ \text{for}\ |\delta-\delta_0|\le\eta,\qquad|c_1Z|\le\tfrac14c_0\ \ \text{at}\ \delta=\delta_0-\eta,\qquad|c_1Z|\ge4c_0\ \ \text{at}\ \delta=\delta_0+\eta .
\label{eq:S-twoterms}
\end{equation}
The series therefore winds zero times around the origin on the inner boundary of the annulus and once on the outer boundary, so $G$ has exactly one simple zero in each cell, with damping angle within $\eta$ of $\delta_0$. Cut each annulus along the ray where $c_1Z/c_0>0$, so that the two leading terms add there; then the series stays above $c_0/2$ on every cell boundary. Since $c_0\ge\pi e^{-h}$, this gives $|G(\lambda)|\ge g_*|\lambda|^\kappa$ with $g_*=p\pi e^{-h}/(2H)$ on every cell boundary, independent of $B$.

\paragraph*{(v) The finite chain.} Three differences separate $\Lambda$ from $G$: the finite products within each excursion, the blocks before the start, and the end of the chain. Equation~\eqref{eq:S-prod}, with $K_0=2\ln2+2|\ln\cos(\delta_0+\eta)|$, bounds them by
\begin{equation}
|\Lambda(\lambda)-G(\lambda)|\ \le\ \frac{11|\lambda|}{h}\left[e^{K_0/h}+\frac{pe^{\kappa H}e^{-h}}{1-e^{-(1-\kappa)H}}\right]+4e^{K_0/h}(1-p)^{B-1}.
\label{eq:S-error}
\end{equation}
The first term is small near $\lambda=0$ and the second in long chains. On every cell between the radii $\lambda_{\rm bot}=b_he^{-H(B-1)}$, with $b_h=(16e^{K_0/h}/g_*)^{1/\kappa}$, and $\lambda_{\rm top}=\min\{h/4,(g_*/4A)^{1/(1-\kappa)}\}$, with $A$ equal to $11/h$ times the bracket, the difference is smaller than $|G|$, so Rouch\'e's theorem gives $\Lambda$ exactly one zero there. Counting the complete cells shows that at least $B-3-[\ln b_h+\ln(1/\lambda_{\rm top})]/H$ blocks keep a rhythm, that is, a loss
\begin{equation}
C\ \le\ 3+\frac{K_0}{\kappa(1-\kappa)sh^2}+C_*\frac{\ln(1/h)}{sh}
\label{eq:S-loss}
\end{equation}
with $C_*$ depending on $s$, $\kappa$, and $\eta$ only. The loss does not grow with $B$, which proves Theorem~\ref{thm:theta}.

\subsection{Finite chains}\label{sec:finitechains}

The chain $\mathcal T(3,0.04,1/2,100)$ of 300 states, with the fastest rate set to one and its eigenvalues checked against Eq.~\eqref{eq:S-renewal} at 40 digits, holds 85 rhythms with $Q>1.4$; a one-way ring of 300 states holds 59. The bound~\eqref{eq:S-loss} is far from tight: for $s=3$, $\kappa=1/2$, $\eta=\pi/12$, and $h=0.04$ it gives no rhythms at $B=100$, whereas the computed spectra lose at most 10 blocks for $h\ge0.025$. The two ends of a chain lose at most $O(h^{-2})$ blocks, so sharing pays off only in long chains. At $Q_0=2$ the best finite chains found hold 78 and 344 rhythms on 400 and 1500 states, against 66 and 249 for standalone clocks (gains of 18\% and 38\%); at $Q_0=10$ no chain tested beats standalone clocks up to 1500 states (\texttt{repro/finite\_d.log}).

\section{Distinguishable rhythms}\label{sec:resolved}

Nested cycles reach the Weyl constant only with blurred rhythms. This section shows that their advantage disappears before their rhythms become distinguishable, that optimized networks find no gain over standalone clocks once all rhythms must be distinguishable, that rhythms on separated time scales are additive, and that a fixed frequency band holds only about $\sqrt d$ distinguishable rhythms.

\subsection{Convention}

Assign rhythm $j$ the interval $[\omega_j-\gamma a_j,\omega_j+\gamma a_j]$ and call two rhythms distinguishable when their intervals are disjoint,
\begin{equation}
|\omega_i-\omega_j|>\gamma(a_i+a_j).
\label{eq:S-resolved}
\end{equation}
The Letter uses $\gamma=1/\sqrt3$: two equal lines $(a^2+(x\pm b)^2)^{-1}$ show a dip at their midpoint exactly when $2b>2a/\sqrt3$. This is a linewidth convention, not a general test of detectability, and the results below hold for every fixed $\gamma>0$. The largest number of pairwise distinguishable rhythms in a given spectrum is found exactly by accepting, again and again, the compatible interval with the earliest right end.

\subsection{The price of distinguishability in nested cycles}\label{sec:price}

In the bulk of a long chain, one block of $s$ states carries one rhythm, whose damping angle $\delta$ and spacing $x=\Delta\omega/\bar a=2\tanh(sh/2)\cot\delta$ to its neighbors, in mean half-widths, depend only on $s$, $h$, and $\kappa$. We computed both from the zeros of $G$ for $s=4$, 8, 16, and 32, with $\kappa$ chosen to minimize $\delta$, and compared the $s$ states per rhythm with the $\pi/\delta-1$ of standalone clocks of equal sharpness. As $h\to0$ the chains approach the bound $\pi/(2\delta)$ of Corollary~\ref{cor:count} while $x\to0$. As $h$ grows the rhythms separate, and the chains lose their advantage at spacings of $0.77$ to $0.92$ half-widths, below the dip threshold $2/\sqrt3\approx1.15$ [Fig.~\ref{fig:sharing}(b); \texttt{repro/sharing\_resolution.py}]. Further out, the damping angle for larger $s$ starts to oscillate with $h$, because the second Fourier term of $G$ becomes comparable to the first two, and the cost stops rising monotonically with the spacing: at $1.28$, $1.10$, and $0.90$ half-widths for $s=8$, 16, and 32, whereas for $s=4$ it rises throughout. Figure~\ref{fig:sharing}(b) therefore ends every curve at the same height, $10\%$ above the cost of standalone clocks, after its crossing and before these oscillations.

\subsection{Optimized networks}\label{sec:optimized}

At $Q_0=1$, standalone clocks of five-state rings with the entry rates of Eq.~\eqref{eq:S-entry} hold $N_{\rm ring}=\lfloor(d-1)/4\rfloor$ distinguishable rhythms, since $\cot(\pi/5)\approx1.376>1$; one more would leave a ring with four states and $Q\le1$. We searched for networks that do better.

\paragraph*{Search.} Two families: all networks on $d$ states, with all $d(d-1)$ rates free, and shared-state networks (one shared state and a path; the shared state can enter every path state, and every path state can return to it), which contain both standalone clocks and nested cycles. Sizes $d=21$, 31, and 41. Targets: the $(N_{\rm ring}+1)$th sharpness with blurring allowed, and the $N_{\rm ring}$th and $(N_{\rm ring}+1)$th sharpness among distinguishable rhythms. Each family, size, and target received 40 starts (24 random unions of driven cycles, 6 standalone clocks, 6 nested cycles, 4 random rate matrices), optimized by Adam on all log-rates for 4000 steps, with derivatives through the eigenvalues. Distinguishability was enforced with a 5\% margin during the search and checked without it; the best runs were confirmed with 40-digit eigenvalues.

\paragraph*{Result.} With blurring allowed, the optimizer reaches $(N_{\rm ring}+1)$th sharpnesses of 1.02, 1.18, and 1.04, that is, 6, 8, and 11 rhythms with $Q>1$: one more than standalone clocks. With distinguishability required, the $(N_{\rm ring}+1)$th sharpness never exceeds 0.99 in any of the 240 runs with this target, so no run holds more distinguishable rhythms with $Q>1$ than standalone clocks (\texttt{data/resolved\_rhythms.json}). Optimization gives lower bounds on the best counts only. The mechanism is visible at branch points: a branch whose rhythm is distinguishable returns out of phase with the others and acts as a leak; only branches with blurred rhythms recombine coherently.

\subsection{Finite sizes and separated time scales}\label{sec:additive}

\paragraph*{An integer effect.} A search at $d=12$ found two distinguishable rhythms with sharpnesses 1.838 and 1.823 (frequency ratio 15.3; \texttt{data/resolved\_windows.json}). Standalone clocks need 13 states for them, because the smaller of two rings sharing one state has at most six states and $Q\le\cot(\pi/6)\approx1.732$. The gain of one state is smaller than the rounding of the two ring sizes (about 6.3 states each), so it does not affect additivity at leading order.

\paragraph*{Separated time scales.} Let each rate scale as $W_{ij}\propto\zeta^{m_{ij}}$ with $\zeta\to0$, and let $N$ rhythms have eigenvalues $\zeta^{\beta_j}(\mu_j+o(1))$ with nonreal $\mu_j$ and distinct $\beta_j$. With $n_j\ge3$ the smallest ring with $\cot(\pi/n_j)\ge|\mathrm{Im}\,\mu_j|/(-\mathrm{Re}\,\mu_j)$,
\begin{equation}
d\ \ge\ 1+\sum_{j=1}^{N}(n_j-1).
\label{eq:S-scales}
\end{equation}
Reduce the network successively on its slow time scales~\cite{coderch1983,rohlicek1988}. A class that carries $\mu_j$ needs at least $n_j$ states by the single-rhythm bound (a transient class after adding one absorbing state), and eliminating it on the next scale removes at least $n_j-1$ states; distinct scales charge these losses only once, and one state survives. Rhythms on widely separated time scales are therefore additive: they need at least the states of standalone clocks. Here the frequency ratios diverge; distinguishability at a finite frequency ratio is a different question.

\subsection{The square-root law}\label{sec:band}

\begin{theorem}[Square-root law]\label{thm:band}
If $N\ge2$ rhythms are pairwise distinguishable in the sense of Eq.~\eqref{eq:S-resolved} and their frequencies span the ratio $R>1$, then
\begin{equation}
d-1\ >\ \frac{\pi\gamma(N-1)^2}{\ln R}.
\label{eq:S-band}
\end{equation}
\end{theorem}

\begin{proof}
Order the rhythms by frequency and put $u_j=\gamma a_j/\omega_j$. Adjacent intervals obey $\omega_j(1+u_j)<\omega_{j+1}(1-u_{j+1})$. Multiplying these inequalities gives $\ln R>\ln(1+u_1)+\sum_{j=2}^{N-1}\ln[(1+u_j)/(1-u_j)]-\ln(1-u_N)$. The two end terms cover at least $\ln[(1+u_\star)/(1-u_\star)]$ for the end $\star\in\{1,N\}$ with the smaller $u_\star$, and $\ln[(1+x)/(1-x)]\ge2x$. So the set $I$ of $N-1$ rhythms (the inner ones and that end) obeys $\sum_{j\in I}a_j/\omega_j<\ln R/(2\gamma)$, and Eq.~\eqref{eq:S-reldamp} gives $\pi(N-1)^2/[2(d-1)]<\ln R/(2\gamma)$.
\end{proof}

\noindent A fixed band therefore holds at most about $\sqrt d$ distinguishable rhythms; a count proportional to $d$ needs a frequency ratio $R$ that grows exponentially with $d$. Ordinary variables are not limited in this way: a block $\left(\begin{smallmatrix}-a&-\omega\\ \omega&-a\end{smallmatrix}\right)$ gives one oscillation with arbitrarily small $a$ on two real variables, so $d$ of them hold $\lfloor d/2\rfloor$ distinguishable rhythms in any band.

\paragraph*{Rings reach the same scaling.} A one-way ring of $n$ states and rate $r$ has $Q=\cot(\pi/n)$ at frequency $r\sin(2\pi/n)$, so $r$ sets the frequency without changing the sharpness. Rings at equal logarithmic spacing are distinguishable in a band of ratio $R$ whenever
\begin{equation}
Q\ >\ \frac{\gamma}{\tanh[\ln R/(2(N-1))]},
\label{eq:S-ringband}
\end{equation}
which needs $d\sim2\pi\gamma N^2/\ln R$ states, twice the bound~\eqref{eq:S-band}; weak links make the union irreducible. On 100 states and one decade ($R=10$, $\gamma=1/\sqrt3$), Eq.~\eqref{eq:S-band} allows at most 12 distinguishable rhythms, eight 12-state rings fit [$\cot(\pi/12)\approx3.73>3.54$], and 100 ordinary variables hold 50.

\section{Drive}\label{sec:driving}

Rhythms need drive. Weak steps cost states: with at most $\varepsilon\,k_BT$ per step, $N$ rhythms sharper than $Q_0$ need $d-1\gtrsim\pi NQ_0/(2\mu)$ states, $\mu=\tanh(\varepsilon/2)$. And every rhythm costs drive: $N$ rhythms sharper than $Q_0>1$ need a total drive above $\pi(N+1)(Q_0-1/Q_0)$, Eq.~\eqref{eq:drive}.

\paragraph*{Setup.} On a link with positive traffic the affinity is $A_e=\ln(\Phi_{ij}/\Phi_{ji})$, the entropy (in units of $k_B$) produced per net jump, and $|J_e|/T_e=\tanh(|A_e|/2)$. Put $\varepsilon=\max_e|A_e|$ and $\mu=\tanh(\varepsilon/2)$; one-way jumps give $\mu=1$, and detailed balance gives $\mu=0$ and no rhythms.

\subsection{The weak-step law}\label{sec:weakstep}

With $b_j=\pi(2j-1)/[2(d-1)]$, every network obeys
\begin{equation}
\sum_{j=1}^{k}\delta_j\ \ge\ \sum_{j=1}^{k}\arctan\frac{\tan b_j}{\mu},
\qquad\text{equivalently}\qquad
\sum_{j=1}^{k}\arctan Q_j\ \le\ \sum_{j=1}^{k}\arctan(\mu\cot b_j).
\label{eq:S-sizediss}
\end{equation}

\begin{proof}
Keep every traffic but amplify every circulation by $1/\mu$~\cite{xu2025}: the flows $\Phi'_{ij}=(T_{ij}+J_{ij}/\mu)/2$ are nonnegative and balanced, so $\Phi'\Pi^{-1}$ is a matrix of jump rates with the same stationary distribution, the same $L$, and $K/\mu$ in place of $K$. Its angles $\hat\alpha_j$ obey $\sum_{j\le m}\hat\alpha_j\ge\sum_{j\le m}b_j=\pi m^2/[2(d-1)]$ by Eq.~\eqref{eq:S-assembly}. In the original network $\cot\alpha_j=\mu\cot\hat\alpha_j$, that is, $\alpha_j=\arctan(\tan\hat\alpha_j/\mu)$, an increasing concave function of $\hat\alpha_j$ that vanishes at zero. Such functions preserve the ordering of partial sums (Tomi\'c--Weyl majorization), and Eq.~\eqref{eq:S-aux} completes the proof.
\end{proof}

\noindent At $\mu=1$ this is Theorem~\ref{thm:counting}. For fixed $\mu$ and large $Q_0$, $N$ rhythms sharper than $Q_0$ need $d-1\gtrsim\pi NQ_0/(2\mu)$. Five rhythms sharper than 10 need at least 80 states for one-way jumps [Fig.~\ref{fig:costs}(a)] and 171 at $\varepsilon=1$; the single-rhythm bound $Q\le \mu\cot(\pi/d)$~\cite{ohga2023} asks for only 32 and 69; standalone clocks use 156 states at $\mu=1$, and five weakly linked rings of 69 states about 345 at $\varepsilon=1$. Long uniform rings, with $Q_m=\mu\cot(\pi m/d)$, attain Eq.~\eqref{eq:S-sizediss} to leading order at every $\varepsilon$; this statement about sums does not settle the best number of rhythms above a fixed sharpness at finite drive.

\subsection{Total drive}\label{sec:ftilde}

The stationary circulation decomposes into directed cycles with circulations $J_c>0$, $\sum_{c\ni e}J_c=|J_e|$~\cite{schnakenberg1976,kalpazidou2006}, for instance by repeatedly subtracting the smallest circulation on a directed cycle~\cite{ahuja1993}. Cycle $c$ has drive $F_c=\sum_{e\in c}|A_e|$; under local detailed balance $k_BTF_c$ is the free energy that drives one turn. The dissipated power is $k_BT\sigma$ with $\sigma=\sum_cJ_cF_c$.

A link on several cycles carries traffic for all of them. Allocate traffic $T_{c,e}\ge J_c$ to each cycle on each of its links, with $\sum_{c\ni e}T_{c,e}\le T_e$. A standalone ring with this traffic and circulation has drive $B_c=\sum_{e\in c}\ln[(T_{c,e}+J_c)/(T_{c,e}-J_c)]$. The total drive is the least total drive of such rings,
\begin{equation}
\Ft=\inf\sum_c\sum_{e\in c}\ln\frac{T_{c,e}+J_c}{T_{c,e}-J_c},
\label{eq:S-ftilde}
\end{equation}
over decompositions and allocations. Proportional allocation, $T_{c,e}=J_cT_e/|J_e|$, gives $B_c=F_c$, so $\Ft\le\sum_cF_c$, with equality when no link lies on two driven cycles.

\subsection{The ladder bound}

A ring of $n_c$ states and drive $B_c$ offers the ladder $r_{c,m}=\tanh[B_c/(2n_c)]\cot(\pi m/n_c)$, $1\le m<n_c/2$, the sharpnesses of its overtones when uniform. With $r_{(j)}$ the $j$th largest rung of all ladders (padded with zeros),
\begin{equation}
\sum_{j=1}^{k}\arctan Q_j\ \le\ \sum_{j=1}^{k}\arctan r_{(j)},\qquad
\sum_{j=1}^{k}\delta_j\ \ge\ \sum_{j=1}^{k}\arctan\frac{2\pi j}{\Ft}.
\label{eq:S-ladder}
\end{equation}
The first bound holds for every decomposition and allocation; the second, its discrete form, follows from $r_{c,m}\le B_c/(2\pi m)$, so that each ring has at most $B_c/(2\pi v)$ rungs above $v$ and $r_{(j)}\le\Ft/(2\pi j)$.

\paragraph*{Why a ring offers this ladder.} For a ring, the spanning-tree polynomial of Eq.~\eqref{eq:S-kirchhoff} telescopes, and its positive ratios $\rho_m$ solve $\sum_{e\in c}\arctan[\tanh(B_{c,e}/2)/\rho_m]=\pi m$, with $B_{c,e}$ the affinity of link $e$ in the ring. The left side is concave in $B_{c,e}$, so Jensen's inequality gives $\rho_m\le r_{c,m}$, with equality for uniform rings.

\paragraph*{Combining the rings.} The rings' circulation forms add up to $K$, and their traffic forms to at most $L$. By the min--max principle, the network has no more ratios above any $v>0$ than the rings together, and Eq.~\eqref{eq:S-aux} gives Eq.~\eqref{eq:S-ladder}. At $k=1$ this is the bound of Ref.~\cite{ohga2023}; for single rings it follows from Ref.~\cite{kolchinsky2026winding}.

\subsection{The drive law}

If $N$ rhythms are sharper than $Q_0>1$, the discrete form requires $\sum_{j\le N}\arctan(2\pi j/\Ft)<N\theta$ with $\theta=\arccot Q_0$. At $\Ft=\pi(N+1)(Q_0-1/Q_0)$ the arguments equal $j\tan(2\theta)/(N+1)$, and concavity of the arctangent on $[0,\tan2\theta]$ makes the sum at least $N\theta$; smaller $\Ft$ give larger sums. Hence
\begin{equation}
\Ft\ >\ \pi(N+1)\bigl(Q_0-Q_0^{-1}\bigr),
\label{eq:S-affinity}
\end{equation}
which is Eq.~\eqref{eq:drive} of the Letter; per rhythm the bound tends to $\pi Q_0$, half the $2\pi Q_0$ that standalone clocks need. For $Q_0=10$ and $N=5$, the discrete form requires $\Ft>188.08$, Eq.~\eqref{eq:S-affinity} gives 186.6 [Fig.~\ref{fig:costs}(a)]. Whether any network reaches the cost $\pi NQ_0$ at high sharpness is open.

\subsection{Why the bound uses sums}\label{sec:certificates}

For a single ring the drive bound holds rhythm by rhythm, $Q_m<F_c/(2\pi m)$; with several cycles it does not. A 27-state network (one driven 20-state cycle with $F_c=2.035$ and a passive side path) has a third rhythm 1.39 times sharper than this law allows, and in a 62-state comb $Q_{30}\ge0.6583$ exceeds the per-index ring value 0.6573. Both are certified with exact rational rates and disjoint inclusion discs of radius ${\lesssim}\,10^{-90}$ (\texttt{repro/verify\_certificates.py}), and both obey Eq.~\eqref{eq:S-ladder}.

\subsection{The drive of shared paths}\label{sec:affcost}

Driving a shared path is expensive. In three chains of nested cycles with affinity 3 on every link, at $Q_0=2.5$, the total drive was 11 to 24 times that of standalone clocks with the same number of rhythms (\texttt{repro/finite\_d.log}).

\subsection{Distinguishable rhythms at finite drive}

Applied to the $N-1$ rhythms selected in the proof of Theorem~\ref{thm:band}, Eqs.~\eqref{eq:S-sizediss} and \eqref{eq:S-ladder} require, at fixed $R$, $\gamma$, and $\mu$ and large $N$, $d-1\gtrsim\pi\gamma(N-1)^2/(\mu\ln R)$ and $\Ft\gtrsim2\pi\gamma(N-1)^2/\ln R$.

\section{Power bounds no rhythm: the parking state}\label{sec:rate}

Take a uniform ring of $n$ states with forward rate $f$ and backward rate $b<f$. Add a parking state that every ring state enters at rate $g$ and that refills each ring state at rate $\chi$. A ring Fourier mode sums to zero, so its eigenvalue $\lambda$ becomes $\lambda-g$, independent of $\chi$: every rhythm keeps its frequency, and its damping grows only by $g$. Choose $g$ much smaller than the damping rates, then let $\chi\to0$. The stationary probability of a ring state is $\chi/(g+n\chi)$, the links of the parking state are balanced, and the dissipated power vanishes,
\begin{equation}
\sigma=\frac{n\chi(f-b)\ln(f/b)}{g+n\chi}\ \longrightarrow\ 0\qquad(\chi\to0).
\end{equation}
With $M$ rings sharing one parking state, the remaining eigenvalues are $0$, $-g$ (multiplicity $M-1$), and $-(g+Mn\chi)$, so the relaxation stays fast while the rings empty. A molecule prepared on the rings shows every rhythm in full while it relaxes. In stationary correlations, by contrast, every rhythm enters with the ring occupation $\chi/(g+n\chi)$ and vanishes together with $\sigma$. No positive lower bound on $\sigma$ in terms of the sharpnesses $Q_1,Q_2,\dots$ can hold; this extends the counterexample of Ref.~\cite{shiraishi2026} to the conjecture of Ref.~\cite{oberreiter2022} from one rhythm to all, by the localization mechanism of Ref.~\cite{gu2026}. As $g\to0$, such rings also approach the ladder of Eq.~\eqref{eq:S-ladder}.

\section{Hearing a hidden clock}\label{sec:inference}

Every fading oscillation in a measured correlation function is a rhythm of the hidden network. Measured sharpnesses therefore certify a minimum number of states and a minimum total drive.

\subsection{Observed rhythms are rhythms of the network}

A stationary correlation has the form $C(\tau)=\bm u^{\mathsf T}e^{W\tau}\bm v$, with Laplace transform $\bm u^{\mathsf T}(\lambda I-W)^{-1}\bm v$. Every pole $\lambda$ is an eigenvalue of $W$: an observable can hide a rhythm but cannot create one, also for nondiagonalizable $W$. Lumped or coarse-grained observables are still matrix elements of $e^{W\tau}$, so coarse graining can only hide rhythms. Count each distinct observed pole pair once.

\subsection{The hearing tests}

Let $k$ rhythms have reliably identified sharpnesses $Q_1\ge\cdots\ge Q_k$, put $\delta_j=\arccot Q_j$ and $S_m=\sum_{j\le m}\delta_j$, and keep the largest of
\begin{equation}
d\ge2k+1,\qquad d\ge\Bigl\lceil\frac{\pi}{\delta_1}\Bigr\rceil,\qquad d\ge1+\Bigl\lceil\frac{\pi m^2}{2S_m}\Bigr\rceil\quad(1\le m\le k).
\label{eq:S-tests}
\end{equation}
The first counts the observed eigenvalues plus the stationary one, the second is the single-rhythm bound, and the third is Theorem~\ref{thm:counting} applied to the $m$ sharpest rhythms. With relative widths in place of angles, the third becomes Eq.~\eqref{eq:hearing} of the Letter, and the second implies $d>\pi Q_1$. The largest drive that violates the discrete form of Eq.~\eqref{eq:S-ladder}, maximized over $m$, is a certified minimum of $\Ft$. When the largest step affinity $\varepsilon$ is known, Eq.~\eqref{eq:S-sizediss} sharpens the state count. These are necessary sizes and drives, not a model, and no separation of the lines is needed once the rhythms are identified.

\paragraph*{Noise and incomplete observation.} Use only statistically supported components and simultaneous lower confidence limits on their sharpnesses, which make the bounds conservative; a damped-sinusoid fit alone does not establish a genuine pole. Measurement noise that is independent of the process and uncorrelated in time leaves every correlation at positive lag unchanged. Detector dynamics, correlated noise, and external forcing need separate treatment.

\subsection{The worked example}

The hidden network of Fig.~\ref{fig:costs}(b) has three uniform rings of 16, 14, and 12 states with forward rates 1, 2.2, and 4.8, backward rates smaller by $e^{-3}$ ($3\,k_B$ of entropy per step), and reversible links of rate 0.02; its total drive is $3\times42=126$. The readout $\lfloor3i/n\rfloor$ of state $i=0,\dots,n-1$ of a ring of $n$ states reports only which third of its ring the process occupies. The correlations $C_{xy}(\tau)$, $x,y\in\{0,1\}$, of 400 stationary trajectories of length 2500 ($2.55\times10^6$ jumps), sampled at $\Delta\tau=0.25$, were fitted jointly by a matrix pencil~\cite{hua1990}. The model order was chosen by twofold cross-validation over trajectories, a rhythm was kept if its pole persisted over seven consecutive orders, and lower confidence limits came from 1000 bootstrap resamples of whole trajectories~\cite{efron1979} at level $0.05/k$ each (95\% jointly). The three lower limits on sharpness are 3.83, 2.99, and 2.95. Equation~\eqref{eq:S-tests} gives $S_3=0.9039$ and $d\ge1+\lceil15.64\rceil=17$, whereas the sharpest rhythm alone gives 13; the relative-width form, Eq.~\eqref{eq:hearing}, gives the same 17. The certified drive is $39.8$ from three rhythms and $24$ from the sharpest alone (\texttt{data/worked\_example.json}). Every Markov model with fewer than 17 states, or with a total drive of at most 39.8, is thus excluded at 95\% confidence.

\subsection{Test of the protocol}\label{sec:protocol}

On 600 simulated data sets from six hidden networks, the certified number of states never exceeded the true one. Point estimates in place of lower confidence limits overstated it in up to 20 of the 50 data sets of one setting (40\%).

\subsection{Stochastic reservoir computers}\label{sec:reservoir}

A weak input changes the rates of a stochastic reservoir~\cite{ehlers2025} in proportion to its strength, $W\to W+V$ per unit input. To first order, its impulse response $\bm f^{\mathsf T}e^{W\tau}V\bm\pi$ is the stationary cross-correlation of the readout $\bm f$ with the observable $g_j=(V\bm\pi)_j/\pi_j$. Its memory rhythms therefore obey the capacity and the drive law: $N$ sharp memory rhythms need the states and the drive these laws require. Input and readout can suppress a rhythm; the laws say nothing about amplitudes or noise. Quantum reservoirs need a separate argument, since a Hamiltonian rings without damping at fixed dimension.

\subsection{Kinesin}

In the six-state kinesin network of Ref.~\cite{liepelt2007}, a fast mechanical transition merges two states on the slow time scale; the remaining five states form two triangles, so each slow rhythm has $Q\le\cot(\pi/3)=1/\sqrt3$~\cite{ohga2023}.

\section{Code and data}\label{sec:code}

The folder \texttt{figsrc} contains the scripts that draw Figs.~\ref{fig:ways}--\ref{fig:costs} and compute their data; its \texttt{README.md} lists the commands. The folder \texttt{repro} contains the reproduction scripts and logs cited above (\texttt{standalone\_clocks.py}, \texttt{sharing\_resolution.py}, \texttt{finite\_d.log}, \texttt{verify\_certificates.py}), and \texttt{data} contains the stored results (\texttt{resolved\_rhythms.json}, \texttt{resolved\_windows.json}, \texttt{worked\_example.json}). The networks behind the figures are:

\medskip
\begin{center}
\renewcommand{\arraystretch}{1.25}
\begin{tabular}{@{}p{0.2\textwidth}p{0.76\textwidth}@{}}
\hline\hline
Panels & Networks\\
\hline
Fig.~\ref{fig:budget}(a) & No network: the modes $\omega=\sqrt{p^2+q^2}$ ($p,q\ge1$) of a square drum of side $\pi$ with fixed edges.\\
Fig.~\ref{fig:budget}(b) & Uniform one-way ring of 20 states with unit rates, started in one state: return probability, occupation of its states (left window), and spectral line of its sharpest rhythm (right window).\\
Fig.~\ref{fig:budget}(c) & 40 random networks for each odd $d$ from 5 to 25 (each link present with probability 0.5, log-normal rates of log-standard deviation 2, plus a random directed cycle); standalone clocks of $m$ states each, $Q=\cot(\pi/m)$ at $k/(d-1)=1/(m-1)$; the one-way ring of 32 states; the chain $\mathcal T(4,0.01,1/2,375)$ of 1500 states, the best chain of Sec.~\ref{sec:finitechains} at $Q_0=2$.\\
Figs.~\ref{fig:sharing}(a), \ref{fig:sharing}(b) & Densities of Sec.~\ref{sec:theta}; bulk rhythms of Sec.~\ref{sec:price} for $s=4$ and 32.\\
Fig.~\ref{fig:costs}(a) & The smallest $d$ allowed by Theorem~\ref{thm:counting} and the discrete form of Eq.~\eqref{eq:S-ladder}, at $Q_0=10$, divided by the cost of standalone clocks, $N(n_0-1)+1$ states ($n_0=32$), and by $2\pi Q_0$ of drive per rhythm, the least drive of one rhythm; single-rhythm bounds $d\ge\max(n_0,2N+1)$ and $\Ft>2\pi Q_0$.\\
Fig.~\ref{fig:costs}(b) & The worked example of Sec.~\ref{sec:inference}.\\
\hline\hline
\end{tabular}
\end{center}

\end{document}